\documentclass[aps,prx,reprint,superscriptaddress]{revtex4-2}

\usepackage{graphicx}
\usepackage{dcolumn}
\usepackage{bm}
\usepackage{amssymb,amstext,amsmath}
\usepackage{xr}

\usepackage{multirow}
\usepackage{bigstrut}
\usepackage{makecell,rotating}
\usepackage{mathrsfs}
\usepackage{booktabs}
\usepackage{threeparttable}
\usepackage{subfigure}
\usepackage{chngpage}
\usepackage{float}
\usepackage{color}
\usepackage{color}
	 \definecolor{darkred}{rgb}{0.75,0,0}
	 \definecolor{darkgreen}{rgb}{0,0.5,0}
	 \definecolor{darkblue}{rgb}{0,0,0.75}
  	 \definecolor{darkorange}{rgb}{1,0.9,0.1}
	 \definecolor{dark}{rgb}{0,0,0}

\begin{document}

\preprint{APS/123-QED}

\title{Doubly heterogeneous networks facilitate the emergence of collective cooperation}
\author{Yao Meng}
\affiliation{%
Center for Systems and Control, College of Engineering, Peking University, Beijing 100871, China}
\author{Sean P. Cornelius}
\affiliation{%
Department of Physics, Toronto Metropolitan University, Toronto ON M5B 2K3, Canada}
\author{Yang-Yu Liu}%
\affiliation{%
Channing Division of Network Medicine, Department of Medicine, Brigham and Women's Hospital and Harvard Medical School, Boston, MA 02115, USA}
\affiliation{%
Center for Artificial Intelligence and Modeling, The Carl R. Woese Institute for Genomic Biology, University of Illinois at Urbana-Champaign, Champaign, IL 61801, USA
}%
\author{Aming Li}
\thanks{amingli@pku.edu.cn}
\affiliation{%
Center for Systems and Control, College of Engineering, Peking University, Beijing 100871, China}
\affiliation{
Center for Multi-Agent Research, Institute for Artificial Intelligence, Peking University, Beijing 100871, China}


\date{\today}

\begin{abstract}

  There is growing recognition that the network structures arising from interactions between different entities in physical, social and biological systems fundamentally alter the evolutionary outcomes.
  Previous paradigm exploring evolutionary game dynamics has assumed that individuals update their strategies at an identical rate, reporting that structurally heterogeneous networks---despite their ubiquity in real systems---generally hinder the emergence of collective cooperation compared to their homogeneous counterparts. Here we solve this paradox by creating a new paradigm where individuals on arbitrary networks are allowed to update strategies at arbitrary, personalized rates, and provide the precise condition under which universal collective cooperation is favored. We find that when individuals' update rates vary inversely with their number of connections, heterogeneous networks actually outperform homogeneous ones in promoting cooperation. This surprising property of such ``doubly heterogeneous" networks cautions against the conventional wisdom that heterogeneous networks are antagonistic to cooperation. We further develop an efficient protocol for optimizing the promotion of cooperation by tuning individuals' update rates in any structure. Our findings highlight that personalized interaction dynamics, beyond structure, in complex networks are fundamental to understanding and promoting collective cooperation.

\end{abstract}

\maketitle

\section{Introduction}
A major achievement in the study of dynamical processes on complex networks has been the realization that the structured systems represented by complex networks significantly alter the evolutionary outcomes of collective dynamics.
In terms of a typical collective dynamic, networks provide an effective way of understanding the emergence of universal collective cooperative behavior---in which individuals pay a cost to confer a benefit to others---in human and animal societies alike \cite{lieberman2005evolutionary,Nowak92Nature,hauert2004spatial,ohtsuki2006simple,allen2017evolutionary,santos2005scale,Sigmund2010,Szolnoki2012,perc2013evolutionary,perc2010coevolutionary,levin2020collective,li2020evolution,szolnoki2009topology}.
Under the prominent metaphor of the prisoner's dilemma \cite{rapoport1965prisoner},
unstructured systems are known to leave no opportunity for the survival of cooperators \cite{hofbauer1998evolutionary,nowak2004emergence}.
In recent decades, researchers have used the language of networks to characterize the connections and interactions between individuals, exploring evolutionary game dynamics in \emph{structured} systems, where individuals update their strategies based on the payoff they obtained from interactions \cite{santos2005scale,ohtsuki2006simple,allen2017evolutionary,maciejewski2014evolutionary,taylor2007evolution,Zhou2021,Su2022}.
The central question is: which network structures promote cooperation, and which hinder it?

In homogeneous networks---where all individuals basically have the same number of connections---a well-known finding is that natural selection favors cooperation if the ratio between the benefit ($b$) provided by a cooperator and the associated cost paid ($c$) exceeds the average number of neighbors $\langle k \rangle$, namely the simple rule $b/c>\langle k \rangle$ \cite{ohtsuki2006simple}.
Yet for ubiquitous heterogeneous structures---wherein different  individuals may have wildly different numbers of connections---both theoretical analysis and numerical simulations suggest that they appear to hinder the emergence of cooperation compared to homogeneous structures \cite{ohtsuki2006simple,allen2017evolutionary,fotouhi2019evolution}.

Despite remarkable advances in our understanding of the emergence of cooperation in networks, existing studies have been based on a key assumption that all individuals update their strategies at the same rate.
For example, a random individual is selected to die and its neighbors spread their strategies by competing for the position (death-birth \cite{ohtsuki2006simple,allen2017evolutionary}).
Alternatively, individuals may uniformly change their strategies by mimicking that of their neighbors (imitation \cite{ohtsuki2006simple}, pairwise comparison \cite{Traulsen2005}).
Yet, real systems are characterized by heterogeneous interaction rhythms among different individuals \cite{barabasi2005origin,karsai2018bursty}.
This prompts us to ask how this \emph{dynamical} heterogeneity might interact with structural heterogeneity to alter the evolution of cooperation.

Here we investigate evolutionary game dynamics under \emph{non-identical} rates of strategy updating.
Specifically, we consider the scenario where individuals are allowed to update their strategies at arbitrary, personalized rates.
We find that non-identical rates of strategy updating can have profound effects on the emergence of cooperation, especially on the ubiquitous heterogeneous structures that are generally reported to be antagonistic to cooperation.

\begin{figure*}[t]
	\centering
	\includegraphics[width= 0.95\linewidth]{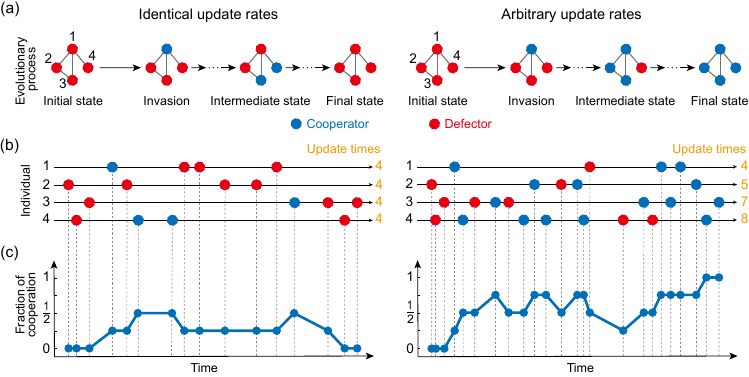}
	\caption{Illustration of the evolutionary process with identical versus arbitrary rates of strategy update.
		The interactions between four individuals are depicted in the example network structure in (a), where individuals play games with their neighbors and gain the corresponding payoffs.
		The evolutionary process starts from a population of full defectors (red), and a cooperator (blue) invades the population via the top site.
		(b) The updating event for each individual occurs as Poisson process.
		We indicate on the timeline when each individual is chosen to update its strategy.
		The color of the dot indicates the strategy after the update, which may be unchanged.
		When individuals' update rates are identical, they will have approximately the same number of strategy updates (numbers in orange, left panel), while for arbitrary update rates, individuals with higher rates will update their strategies more often (right panel).
		The change in the fraction of cooperation throughout the game is illustrated in (c), and the evolutionary process ends when the population reaches a state of either full defection (left panel) or full cooperation (right panel).
		\label{Fig1}
	}
\end{figure*}

\begin{figure*}[t]
	\centering
	\includegraphics[width=0.95\linewidth]{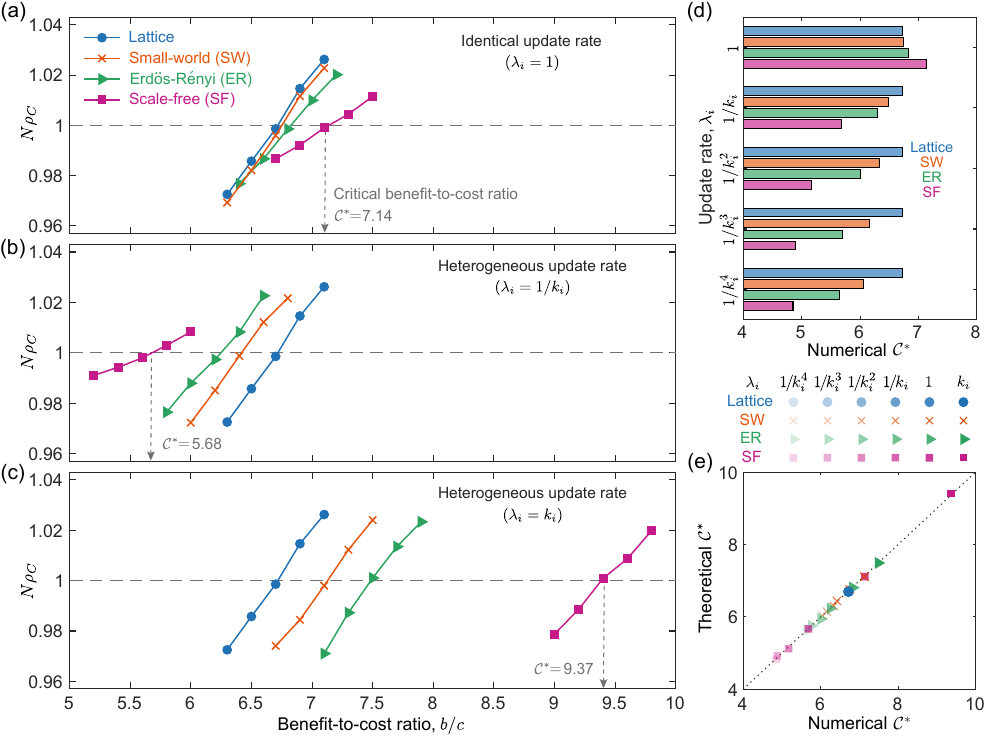}
	\caption{Effect of heterogeneous strategy update rates on the emergence of cooperation.
		We show the fixation probability of cooperation ($\rho_C$) as a function of the benefit-to-cost ratio ($b/c$) over different settings of the update rate ($\lambda_i$) of individual $i$, namely identical ($\lambda_i = 1$ for every individual in (a)) and heterogeneous ($\lambda_i = 1/k_i$ in (b) where $k_i$ is the number of neighbors of $i$, $\lambda_i = k_i$ in (c)) on random regular (RR), Erd\"{o}s-R\'enyi \cite{Erdos1959} (ER), small-world \cite{watts1998collective} (SW) and scale-free \cite{Barabasi1999a} (SF) networks, respectively.
		The critical benefit-to-cost ratio $\mathcal{C}^*$ above which the cooperation is favored for each network occurs when the corresponding curve intersects the horizontal line representing the neutral-drift case ($\rho_C =1/N$).
		$\mathcal{C}^*$ for the scale-free case (purple) is marked.
		We demonstrate that the trend of $\mathcal{C}^*$ reverses when the update rate varies inversely with $k_i$ in (b), presenting the advantage of SF networks on favoring cooperation.
		(d) The ordering of $\mathcal{C}^*$ for the four networks considered holds with $\lambda_i=1/k_i^{\gamma}$ ($\gamma=1,2,3,4$).
		Here we also show that SF networks are the most amenable to cooperation at non-identical update rates compared with other networks.
		(e) Simulation results on $\mathcal{C}^*$ in (a)--(d) are in good agreement with our theoretical calculations shown in Eq.~(\ref{exact_bcr}).
		Numerical values of $\rho_C$ are obtained from the fraction of simulations in which the population reaches full cooperation out of $10^7$ independent realizations on networks of $98$ nodes for lattice and $100$ for other networks with an average degree $\langle k \rangle=6$, and $\delta = 0.01$.
		\label{Fig2}
	}
\end{figure*}

\section{Model}
We consider evolutionary game dynamics on a structured population of $N$ players, whose interactions are represented by a network. 
At any given time, the state of each node (player) is characterized by a strategy of either cooperation ($\text{C}$) or defection ($\text{D}$) [Fig.~\ref{Fig1}(a)].
In each round of the game, every node $i$ plays the game pairwise with its immediate $k_i$ neighbors. Specifically, cooperators pay a cost $c$ to provide a benefit $b$ to each of their neighbors, while defectors pay nothing, and thus provide no benefit. In this way, each node $i$ gains an average payoff $f_i$, corresponding to the average benefits received (from neighboring cooperators) minus its cost.

Traditionally, individuals are assumed to update their strategies following independent Poisson processes with identical rates.
Here we depart from this practice: allowing each individual $i$ to update its strategy with arbitrary rate $\lambda_i$ [Fig.~\ref{Fig1}(b)].
When an individual is chosen for an update, it does so by copying the strategy of one of its neighbors $j$, with probability proportional to the fitness of $j$, generally defined as $F_j = 1+\delta f_j$, where $\delta>0$ captures the intensity of selection \cite{ohtsuki2006simple,allen2017evolutionary}.
Note that for large selection intensity, we still lack theoretical methods to analyze the corresponding nonlinear dynamics \cite{ibsen2015computational}.
Thus, in order to systematically uncover the effects of heterogeneous update rates on the fate of cooperators, here we focus on the canonical case of weak selection.

To quantify the ability of cooperation to proliferate, we initialize our simulations with a single cooperator placed uniformly at random in a population among $N-1$ defectors.
The evolutionary game ends when a state with either all cooperators or all defectors is reached [Fig.~\ref{Fig1}(c)].
We define the \emph{fixation probability} of cooperation ($\rho_C$) as the probability of reaching the state of full cooperation over many realizations of this process.
We can analogously define a probability $\rho_D$ of reaching a full-defection state starting from a single defector planted of $N-1$ cooperators.
Our interest in this study is the condition under which cooperation is favored to replace defection than vice versa \cite{nowak2004emergence,ohtsuki2006simple,allen2017evolutionary}, namely $\rho_C>\rho_D$.
This condition is equivalent to $\rho_C>1/N$ (see Supplemental Material Sec.~I \cite{Supplemental}), namely that selection favors the emergence of cooperation relative to the neutral drift ($\delta=0$), in which neither cooperation nor defection is favored ($\rho_C=\rho_D=1/N$).

\begin{figure*}[t]
	\centering
	\includegraphics[width=0.95\linewidth]{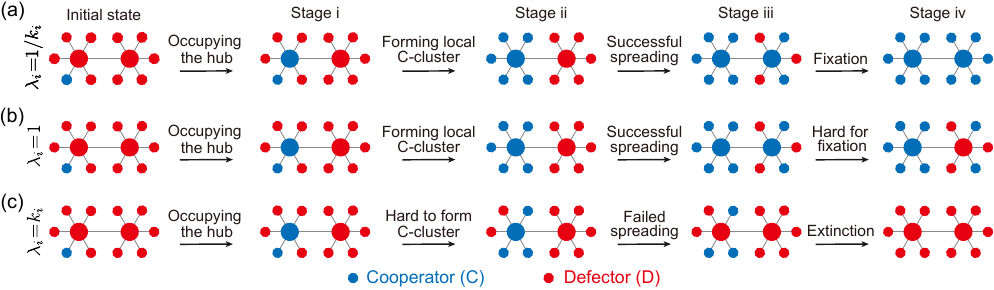}
	\caption{Illustration of the role of hubs on the evolution of cooperation on a double star structure.
		(a) The hubs, two centers of the double star structure for example, have low update rates when $\lambda_i=1/k_i$ ($k_i$ is the number of connections for each node), which facilitates the formation of local cluster of cooperation (blue dot, stage ii) once it is occupied by a cooperator (stage i).
		Likewise, once the left hub spreads cooperation to the right hub (stage iii), the remaining nodes are quickly driven to cooperators (stage iv).
		(b) When the update rates are identical ($\lambda_i=1$), the hubs have many opportunities to change their strategy to defection before all neighbors become cooperators (stage iv), making the fixation of cooperation less likely.
		(c) The hub switches its strategy quite frequently when $\lambda_i=k_i$, which makes it hard to form even the left C-cluster (stage ii), to say nothing of spreading cooperation to the right center.}
		\label{Fig3-doublestar}
\end{figure*}

\section{Results}
\subsection{Evolutionary game dynamics on complex networks}
First, we explore how the heterogeneous strategy updating affects the fate of cooperators on four commonly-studied population structures: lattice, small-world, Erd\"{o}s-R\'enyi, and scale-free networks [Fig.~\ref{Fig2}].
Under the traditional scenario of identical update rates ($\lambda_i=1$ for all $i$), scale-free networks demand the largest critical benefit-to-cost ratio  $\mathcal{C}^*$, above which cooperation is favored among all the four structures, and the lattice structure the smallest [Fig.~\ref{Fig2}(a)],  consistent with previous findings \cite{ohtsuki2006simple,allen2017evolutionary}.
But surprisingly, when a node's update rate varies inversely with its number of neighbors ($\lambda_i=1/k_i$),  we find that this trend is reversed [Fig.~\ref{Fig2}(b)].
Here, scale-free network becomes the most amenable to cooperation, and lattice the least.
Interestingly, we find that heterogeneous update rates can even improve upon the canonical threshold $b/c>\langle k \rangle$ (namely, $\mathcal{C}^*=\langle k \rangle$) for homogeneous populations \cite{ohtsuki2006simple}, allowing cooperation to emerge even when $b/c<\langle k \rangle$ (namely, $\mathcal{C}^*<\langle k \rangle$).
Furthermore, we find that this pattern is strengthened when the update rate is inversely proportional to higher powers of $k_i$ [Fig.~\ref{Fig2}(d)]. 
In contrast, when $\lambda_i$ is positively related to $k_i$, the ordering of $\mathcal{C}^*$ over different structures matches the identical-rate case, but with the inhibition of cooperation fixation by heterogeneous networks amplified [Fig.~\ref{Fig2}(c)].

We further shed light on our numerical findings by deriving a closed-form expression for the critical benefit-to-cost ratio $\mathcal{C}^*$ as a function of the network structure (see Appendix A)
\begin{equation}
	\mathcal{C}^* = \frac{\sum_{i,j} k_i p_{ij}^{(2)}\eta_{ij}}{\sum_{i,j}k_i p_{ij}^{(3)}\eta_{ij}-\sum_{i,j}k_i p_{ij}\eta_{ij}}.
	\label{exact_bcr}
\end{equation}
Here, $ k_i= \sum_{j} e_{ij}$ defines the number of neighbors (degree) of individual $i$, and $e_{ij}=e_{ji}=1$ indicates that there is an edge between nodes $i$ and $j$ ($e_{ij}=e_{ji}=0$ otherwise).
The probability of a $1$-step ($n$-step) random walk from $i$ to $j$ is denoted by $p_{ij}$ ($p_{ij}^{(n)}$),
and $\eta_{ij}$ is  the \emph{coalescence time} \cite{cox1989coalescing}---the expected time for two random walks starting from nodes $i$ and $j$ to meet at a common node.
As shown in Fig.~\ref{Fig2}(e), all numerical results in Figs.~\ref{Fig2}(a)--\ref{Fig2}(d) are in good agreement with the theoretical prediction of Eq.~(\ref{exact_bcr}).

\begin{figure*}[t]
	\centering
	\includegraphics[width=0.95\linewidth]{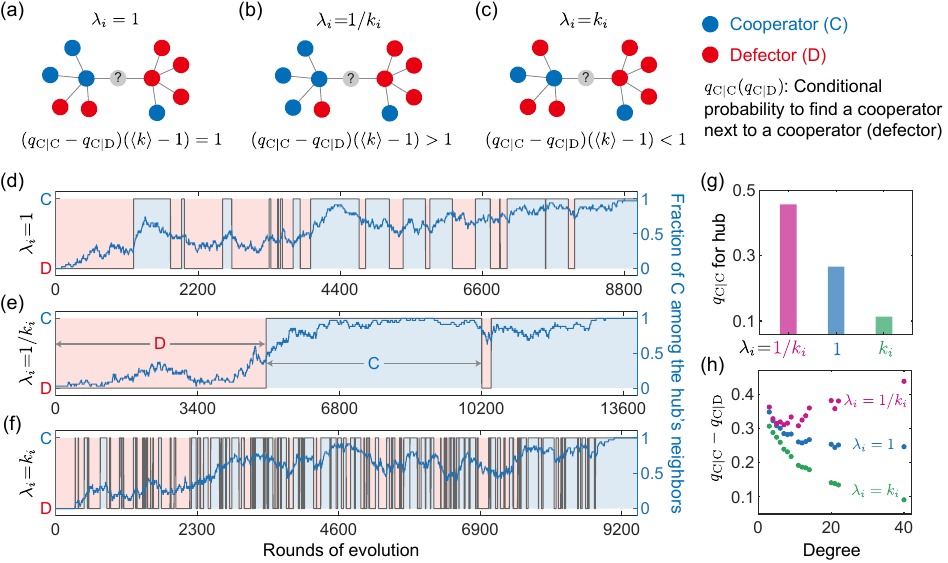}
	\caption{Mechanism for promoting collective cooperation with infrequent strategy updates of hubs.
		(a)--(c) Illustration on the scenario where a cooperator (blue dot) and a defector (red dot) compete to spread their strategy to the individual (grey dot) selected for strategy update under different update rates $\lambda_i$.
		Since behavior dispersal occurs in the neighborhood, the cooperator obtains on average $b(q_{\text{C}|\text{C}}-q_{\text{C}|\text{D}})(\langle k \rangle -1)/\langle k \rangle -c$  more payoff than the defector (Eq.~(\ref{intuition})),
		and the cooperator is favored when the above expression is positive.  
		(a) For identical updating ($\lambda_i=1$), the cooperator has one more cooperative neighbor than the defector, therefore it receives $b/\langle k \rangle$ more benefit than the defector at a cost of $c$.
		(b) When $\lambda_i=1/k_i$, the net benefit of the cooperator relative to the defector exceeds $b/\langle k \rangle$ on heterogeneous networks because the fraction of cooperative neighbors of the cooperator further increases compared to the defector, offering the cooperator a higher chance for dispersal.
		(c) We show that the fast strategy update of hubs ($\lambda_i=k_i$) on heterogeneous networks reduces the number of cooperative neighbors of the cooperator, which exceeds that of the defector by less than one.
		This lower the benefit of the cooperator and reduce the chance to win the empty site.
		(d)--(f) We further compare the state of the hub (grey lines) and the fraction of cooperation among its neighbors (blue lines) of a scale-free network  with different settings of update rates.
		Generally, the hub imitates one of its cooperative neighbor and keeps cooperation for several rounds (light blue shaded region) before switching to defection (light red shaded region) in (d).
		Statistically, we count the fraction of cooperators in the neighborhood of a cooperative hub ($q_{\text{C}|\text{C}}$ for the hub) throughout evolutionary process in (g), and $q_{\text{C}|\text{C}}-q_{\text{C}|\text{D}}$ for nodes with different degree in (h).
		Numerical calculations confirm the mechanism we present in (a)--(c).
		Here, we use the same network parameters as Fig.~\ref{Fig2}.}
		\label{Fig3}
\end{figure*}

\subsection{Role of network hubs}
To intuitively understand why heterogeneous update rates can improve the fixation of cooperation in heterogeneous networks, we first consider how the evolutionary dynamics play out on a simple double star structure [Fig.~\ref{Fig3-doublestar}].
When the fixation of cooperation occurs in this highly heterogeneous structure, it usually does so in four stages: (i) occupation of one of the hubs; (ii) formation of a stable cluster of cooperators among that hub and its neighbors; (iii) occupation of the other hub; and finally (iv) spread to the remaining nodes.
As such, the ultimate triumph of cooperators can be thwarted if a hub imitates defection from even one of its (many) neighbors before stages (ii) and (iv) are complete [Fig.~\ref{Fig3-doublestar}(c)].
There are ample opportunities for this to occur under the traditional setting of identical update rates ($\lambda_i=1$), as illustrated in Fig.~\ref{Fig3-doublestar}(b).
When $\lambda_i=1/k_i$ however [Fig.~\ref{Fig3-doublestar}(a)], hubs update relatively infrequently.
As such, once a hub becomes a cooperator, it is effectively ``locked in", giving time for its strategy to spread to the hub's neighbors.
By the same logic, the preferential updating of hubs ($\lambda_i=k_i$) usually leads to the extinction of cooperation, as the formation of stable clusters of cooperators and the spread of cooperation is even harder than the traditional scenario of identical updating [Fig.~\ref{Fig3-doublestar}(c)].

In Fig.~\ref{Fig3}, we illustrate the fundamental mechanism explaining why infrequent updates of hubs can facilitate cooperation.
If an individual [grey node in Fig.~\ref{Fig3}(a)] decides to update its strategy, it will imitate the strategy of its neighbors according to their payoffs. 
The neighboring cooperator obtains an average payoff $P_\text{C}=b q_{\text{C}|\text{C}}(\langle k \rangle-1)/\langle k \rangle - c$ and the neighboring defector obtains  $P_\text{D}=b q_{\text{C}|\text{D}}(\langle k \rangle-1)/\langle k \rangle$, where $q_{\text{C}|\text{C}}$ ($q_{\text{C}|\text{D}}$) represents the conditional probability to find a cooperative neighbor for a given cooperator (defector).
The contribution to the neighboring cooperator and defector from the updating individual is excluded since they are equal.
Thus the cooperator is favored compared to the defector to disperse its strategy if $P_\text{C}>P_\text{D}$, namely 
\begin{equation}
	b(q_{\text{C}|\text{C}}-q_{\text{C}|\text{D}})(\langle k \rangle -1)/\langle k \rangle-c>0.
	\label{intuition}
\end{equation}
For the canonical setting with identical update rates ($\lambda_i=1$), we know $(q_{\text{C}|\text{C}}-q_{\text{C}|\text{D}})(\langle k \rangle-1)=1$ according to pair approximation (see Supplemental Material Sec.~II \cite{Supplemental}), namely a cooperator has on average one more cooperative neighbor than a defector [Fig.~\ref{Fig3}(a)].
This leads to the conclusion that cooperation is favored when $b/c>\langle k \rangle$ (namely, $\mathcal{C}^*=\langle k \rangle$), which also degenerates to the simple rule \cite{ohtsuki2006simple} for homogeneous networks where $k_i = \langle k \rangle$.

\begin{figure*}[t]
	\centering
	\includegraphics[width=0.95\textwidth]{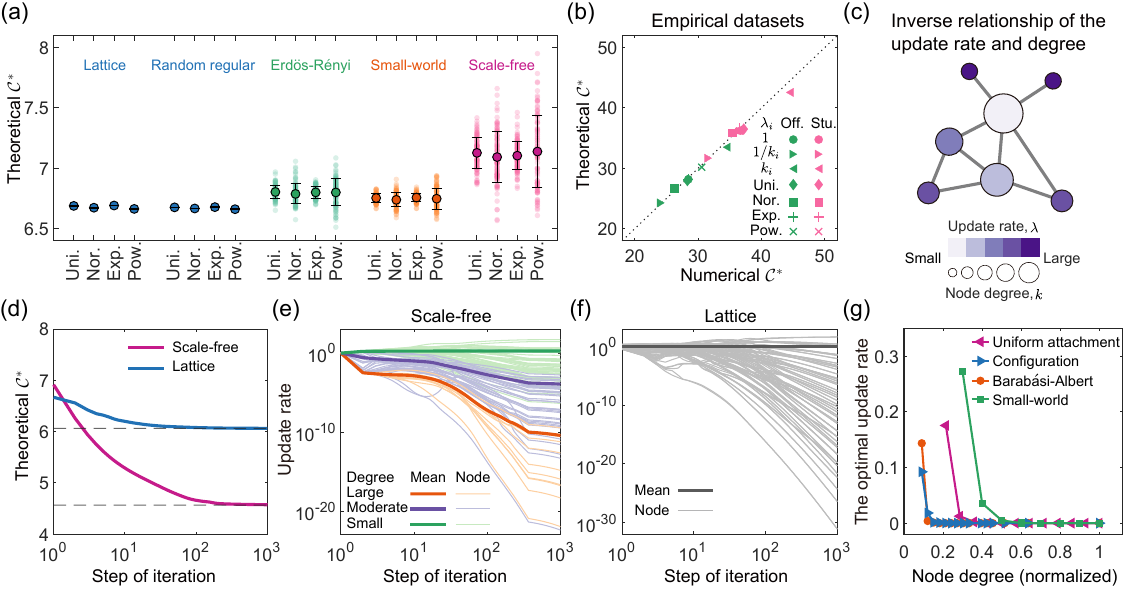}
	\caption{Designing the optimal update rates to promote cooperation on heterogeneous networks.
	(a) Illustration of $\mathcal{C^*}$ above which cooperation is favored under uniform (Uni.), normal (Nor.), exponential (Exp.) and power-law (Pow.) distributions of update rates on different structures of networks.
	Here each dot corresponds to a sample, and the error bars are plotted over $100$ samples.
	The consistent theoretical evidences and details are given in Fig.~S1 \cite{Supplemental}.
	(b) $\mathcal{C}^*$ obtained from our theory (Eq.~(\ref{bcr_approx})) with various update rate ($\lambda_i$) configurations (different markers) are well-matched with the numerical simulations on empirical networks corresponding to face-to-face contacts in an office building \cite{office2013} (Off.) and a high school \cite{student2012} (Stu.).
	Based on the analytical condition given in Eq.~(\ref{bcr_approx_heter_large}), we seek to reduce $\mathcal{C}^*$ on large heterogeneous structures, specifically by letting nodes' update rates vary inversely to their degree $k_i$ as shown in (c), where the size (color) of nodes captures the magnitude of $k_i$ ($\lambda_i$).
	(d) We present the convergence of the objective function $\mathcal{C}^*$ for a scale-free network (purple) and a lattice (blue) over $10^3$ iterations of our optimization protocol.
	(e) We show the corresponding evolution of the (tunable) $\lambda_i$ for all nodes, which are divided into three categories (large, moderate and small) based on the range of degrees in the scale-free network.
	The mean update rate among individuals in each category is shown with the thicker line. 
	We see that the optimal update rates tend to decrease for large nodes (orange) and generally increase for small nodes (green).
	(f) For the lattice, the optimal update rate also presents the deviations from the identical rate.
	Beyond presenting the detailed process for optimizing $\mathcal{C}^*$ in panels (d)--(f), we show the final $\lambda_i$ compared to the nodes' degree for scale-free networks (generated by the configuration model \cite{molloy1995critical}, Barab\'asi-Albert model \cite{Barabasi1999a}), small-world network \cite{watts1998collective} (rewiring probability $0.7$) and networks constructed from a uniform attachment model \cite{barabasi1999mean} in (g), where we normalize the optimal update rate and the node degree.
	We again observe an inverse relationship between the final update rates and the corresponding nodes' degree, consistent with our rule shown in (c).
	Here we use the same network parameters as Fig.~\ref{Fig2}.
	}
	\label{Fig4}
\end{figure*}
Next we show how heterogeneous update rate alters the local dispersal of cooperation on heterogeneous networks.
When $\lambda_i=1/k_i$, we find that $(q_{\text{C}|\text{C}}-q_{\text{C}|\text{D}})(\langle k \rangle-1)>1$ (see Supplemental Material Sec.~II \cite{Supplemental}), indicating that the number of cooperative neighbors of a cooperator exceeds that of a defector by more than one [Fig.~\ref{Fig3}(b)].
This implies that the net payoff of cooperators relative to defectors is further increased, giving cooperators more advantage in competition and dispersal.
Therefore, the critical ratio for $\lambda_i=1/k_i$ is smaller than the average degree $\langle k \rangle$ for a wide range of heterogeneous networks ($\mathcal{C}^*<\langle k \rangle$).
In contrast, when $\lambda_i=k_i$, the hubs update frequently and $(q_{\text{C}|\text{C}}-q_{\text{C}|\text{D}})(\langle k \rangle-1)<1$ (see Supplemental Material Sec.~II \cite{Supplemental}), indicating that on average, the number of cooperative neighbors of a cooperator exceeds that of a defector by less than one [Fig.~\ref{Fig3}(c)].
This leads to a larger critical ratio ($\mathcal{C}^*>\langle k \rangle$) for promoting cooperation compared to the scenario with identical update rates shown in Fig.~\ref{Fig3}(a).

We have numerically confirmed the above mechanism on larger scale-free networks.
Figures~\ref{Fig3}(d)--\ref{Fig3}(f) show the state of the hub, and the fraction of cooperators among the hub's neighbors over the course of the game dynamics.
For $\lambda_i = 1/k_i$, we observe long-lasting periods of cooperation on the hub [Fig.~\ref{Fig3}(e)], with infrequent strategy switches from cooperation to defection, which results in the highest $q_{\text{C}|\text{C}}$ for the hub [Fig.~\ref{Fig3}(g)] and in turn the highest $q_{\text{C}|\text{C}}-q_{\text{C}|\text{D}}$ over all nodes with different degrees compared to other settings [Fig.~\ref{Fig3}(h)].
In contrast, fast-updating hubs ($\lambda_i=k_i$) have the lowest average fraction of cooperators among their neighbors [Fig.~\ref{Fig3}(g)], leading to a low fraction of cooperative neighbors for the cooperators relative to defectors over the whole network [Fig.~\ref{Fig3}(h)].
This confirms that degree-inverse update rates promote cooperation on heterogeneous networks because a hub with a low update rate is more conducive to driving its neighbors to cooperation, which further enhances the local dispersal of cooperation among nodes with different degrees.

\begin{table*}[bt]
    \centering
    \caption{Critical benefit-to-cost ratio $\mathcal{C^*}$ for the fixation of cooperation under different update rates and network structures. For homogeneous networks, $\mathcal{C^*}$ is always equal to the average degree $\langle k \rangle$, irrespective of identical and heterogeneous update rates (see Fig.~\ref{Fig4}(a) for numerical calculations).
    While heterogeneous networks can present quantitatively different values of $\mathcal{C^*}$ under different update rates (Eq.~(\ref{bcr_approx_heter_large})), being determined by the relationship between $k_i$ and $k_j$, $\lambda_i$ and $\lambda_j$ [Figs.~\ref{Fig2}(b),~\ref{Fig2}(c) and~\ref{Fig4}(c)].
    $\lambda_i$ is the update rate for individual $i$ with the number of neighbors $k_i$.}
     	 \setlength{\tabcolsep}{13.2mm}
     	\begin{threeparttable}
    \begin{tabular}{ccc}
		\midrule
		\specialrule{0em}{0.5pt}{1pt}
		\midrule
        Network  & Strategy update rate ($\lambda_i$)   & Critical ratio ($\mathcal{C^*}$)  \\
        \midrule
        Homogeneous  & Identical ($\lambda_i=1$) or heterogeneous  &  $\approx  \langle k \rangle $ \\
        \midrule
        \multirow{3}{*}{Heterogeneous} & Identical ($\lambda_i=1$)  &  $\approx \langle k \rangle$ \\
         &  Heterogeneous, $(k_i-k_j)(\lambda_i-\lambda_j)>0$& $>\langle k \rangle$ \\
         &Heterogeneous, $(k_i-k_j)(\lambda_i-\lambda_j)<0$  &  $<\langle k \rangle$\\
		\midrule
		\specialrule{0em}{0.5pt}{1pt}
		\midrule
    \end{tabular}
     	\end{threeparttable}
	\end{table*}

\subsection{Theoretical analyses}
We next explore how different distributions of $\lambda_i$ affect $\mathcal{C}^*$ over four different synthetic networks: random regular, Erd\"{o}s-R\'enyi, small-world, and scale-free.
For a given network structure, we theoretically predict $\mathcal{C}^*$ via Eq.~(\ref{exact_bcr}) for uniform, normal, exponential and power-law distributions of the update rate.
We find that the critical threshold of a typical homogeneous network---such as a lattice or random regular network---is almost unaffected by the choice of update rate distribution [Fig.~\ref{Fig4}(a)].
In contrast, heterogeneous structures are quite sensitive, with scale-free networks presenting the most drastic variations in $\mathcal{C}^*$ among the different update-rate distributions we consider.
This malleability of $\mathcal{C}^*$ in heterogeneous networks suggests the possibility of deliberately tuning the update rates to lower the barrier for the emergence of cooperation in a particular network.
But to put this into practice, we must first overcome a computational hurdle.

In order to calculate $\mathcal{C}^*$ using Eq.~(\ref{exact_bcr}), one needs to solve a system of $N(N-1)/2$ linear equations for the recurrence relations between the $\eta_{ij}$ (see Eq.~(\ref{eta_ij}) in Appendix A). Unfortunately, this requires an overall complexity of $\mathcal{O} (N^6)$,  rendering the problem intractable for large networks.
To circumvent this, we offer an efficient approximation $\mathcal{C}^*$ as
\begin{equation}
	\mathcal{C}^* \approx \frac{  N \langle k \rangle^2 \zeta/{\langle k^2\rangle}-1 + \Delta_{\lambda^{(1)}}+\Delta_{\widetilde{\eta}_n}}
	{ N\langle k \rangle \zeta/ {\langle k^2\rangle} -1+ \Delta_{\lambda^{(2)}}+\Delta_{\widetilde{\eta}_d}}.
	\label{bcr_approx}
\end{equation}
This expression obviates the need to solve large systems of linear equations and reduces the computational complexity to $\mathcal{O} (N^3)$.
Here $\langle k^2\rangle$ is the second moment of the degree distribution.
We have $\zeta=  \sum_{i,j } \frac{k_i k_j \Lambda}{NK^2(\lambda_i+\lambda_j)}$, where $\Lambda=\sum_{i} \lambda_i$ defines the total rate of update events and $K = \sum_{i} k_i$ is the summation of all nodes' degrees.
Finally, $\Delta_{\lambda^{(1)}}$, $\Delta_{\lambda^{(2)}}$, $\Delta_{\widetilde{\eta}_n}$ and $\Delta_{\widetilde{\eta}_d}$ are constants related to the heterogeneity of update rates and coalescence times, the expressions for which are given in Appendix B.
When the update rates are identical, we have $\Delta_{\lambda^{(1)}}=\Delta_{\lambda^{(2)}}=\Delta_{\widetilde{\eta}_n}=\Delta_{\widetilde{\eta}_d}=0$, and Eq.~(\ref{bcr_approx}) recovers the previous results \cite{fotouhi2019evolution,ohtsuki2006simple}.

Figure~\ref{Fig4}(b) compares the value of $\mathcal{C^*}$ predicted by the approximation in Eq.~(\ref{bcr_approx}) with that of numerical simulation on two empirical social networks \cite{office2013,student2012}. 
We see that our approximation is remarkably accurate in both networks, regardless of the distribution of the update rates.
Moreover, Eq.~(\ref{bcr_approx}) offers intuition behind our previous observation that homogeneous structures are robust to different update rates [Fig.~\ref{Fig4}(a)].
The high symmetry present in these networks means that heterogeneous update rates affect only a limited number of nodes.
For such networks, we have $\Delta_{\widetilde{\eta}_{n}} \approx \Delta_{\widetilde{\eta}_{d}} \approx0$, meaning that $\mathcal{C}^*\rightarrow \langle k \rangle$ in the limit of large $N$.
This coincides with the classical result ($\mathcal{C}^*=\langle k \rangle$) \cite{ohtsuki2006simple} regardless of the distribution of update rates.

\subsection{A simple condition for the emergence of cooperation}
Starting from Eq.~(\ref{bcr_approx}) (see Appendix B), we have the critical benefit-to-cost ratio for large heterogeneous networks
\begin{equation}
	\mathcal{C}^* \approx \langle k \rangle+\frac{\langle k \rangle^2\langle k^2\rangle\Delta_{\widetilde{\eta}^{(\infty)}}}
	{\langle k \rangle^3 \zeta +(\langle k \rangle^3-\langle k \rangle\langle k^2\rangle-\langle k^2\rangle)\Delta_{\widetilde{\eta}^{(\infty)}}},\label{bcr_approx_heter_large} 
\end{equation}
where $\langle k \rangle$ is the average degree and $\Delta_{\widetilde{\eta}^{(\infty)}}\approx\frac{\overline{\eta}}{K^{2}}\sum_{i<j}(k_i-k_j)(\lambda_i-\lambda_j)e_{ij}/(\lambda_i+\lambda_j)$.
Note that $\Delta_{\widetilde{\eta}^{(\infty)}}<0$ when any pair of nodes $i$ and $j$ satisfies the rule $(k_i-k_j)(\lambda_i-\lambda_j)<0$.
When the update rates are identical, we have $\Delta_{\widetilde{\eta}^{(\infty)}} = 0$ and hence $\mathcal{C}^* \approx \langle k \rangle$ as expected.
In contrast, $\mathcal{C}^*$ is smaller (larger) than $\langle k \rangle$ when $\Delta_{\widetilde{\eta}^{(\infty)}}<0$ ($\Delta_{\widetilde{\eta}^{(\infty)}}>0$) (see Supplemental Material Sec.~III \cite{Supplemental}).
Table 1 summarizes the values of $\mathcal{C}^*$ predicted by Eq.~(\ref{bcr_approx_heter_large}) for the combinations of network structure/update-rate settings.

Taken together, we have theoretically motivated an efficient rule of thumb for lowering the threshold for the emergence of cooperation on large heterogeneous structures.
Put simply,  the order of nodes' update rates (for example, $\lambda_i>\lambda_j$) should be reversed from the order of the nodes' degrees (for example, $k_i<k_j$).
That is, nodes with larger degree should have smaller update rates and vice versa, as is demonstrated in Fig.~\ref{Fig4}(c).
A simple but general realization of this rule is $\lambda_i=1/k_i^{\gamma}~(\gamma>0)$ which we study numerically in Fig.~\ref{Fig2}(d) for different values of $\gamma$. 
This rule can achieve a lower critical ratio $\mathcal{C}^*$ than identical update rates ($\gamma = 0$) on both synthetic heterogeneous [Fig.~\ref{Fig2}(b)] and empirical networks ($\gamma=1$) [Table~S1, Figs.~\ref{Fig4}(b),~S2 and~S3 \cite{Supplemental}].
Meanwhile, the contrary configuration of $\lambda_i=k_i^{\gamma}$ leads to increases in $\mathcal{C}^*$ on heterogeneous networks [Figs.~\ref{Fig2}(c),~\ref{Fig4}(b),~S4 and~S5 \cite{Supplemental}].

\subsection{The optimal update rate on any network}
Can we improve upon the simple heuristic to favor cooperation in the previous section? 
Specifically, can we find an \emph{optimal} set of node update rates for a given network?
To answer this question, we employ a protocol based on the RMSprop \cite{tieleman2012lecture} algorithm to search for a set of $\lambda_i$ that minimizes $\mathcal{C}^*$, via iterative gradient descent (see Supplemental Material Sec.~IV and Fig.~S6 \cite{Supplemental}).
Consistent with our rule, Fig.~\ref{Fig4}(d) shows scale-free networks are more flexible and attain a much smaller threshold at its optimal rate than lattice.
Moreover, the update rates of higher-degree nodes tend to decrease during the optimization process, while those of smaller-degree nodes increase [Figs.~\ref{Fig4}(e) and~S7 \cite{Supplemental}].
Interestingly, we find that even on homogeneous structures such like lattices, 
a policy of identical update rates is not the best choice for promoting cooperation.
Indeed, the final update rates deviate significantly from the initial conditions [Figs.~\ref{Fig4}(f) and~S8 \cite{Supplemental}].
Figure~\ref{Fig4}(g) shows that the optimal update rates for different network structures are consistent with our rules shown in Fig.~\ref{Fig4}(c)---namely that a node $i$'s update rate $\lambda_i$ should vary inversely with its degree $k_i$.

\section{Discussion and outlook}

Our findings reconcile the past conflicting results on how heterogeneous networks affect the evolution of cooperation.
Studies that initialize evolutionary game dynamics with an equal number of cooperators and defectors have found that scale-free networks actually outperform homogeneous networks in promoting the evolution of cooperation, as measured by the average fraction of cooperators \cite{santos2005scale}.
But from the perspective of fixation probability, heterogeneous structures impose a higher benefit-to-cost threshold for a single cooperator to take over a population of defectors, at least when all update rates are identical \cite{ohtsuki2006simple,allen2017evolutionary,fotouhi2019evolution}.
This predicts that heterogeneous network structures, despite their ubiquity in physical and social systems, tend to hinder the emergence of collective behavior.
By relaxing this assumption and allowing nodes to update their strategies at non-identical rates, we have shown that scale-free networks can in fact facilitate the fixation of cooperation.
As such, degree-heterogeneous networks orchestrated by personalized update rates can be unambiguously conducive to cooperation, provided they are ``doubly heterogeneous''---that is, also heterogeneous in update rate.
Taken together, we argue that personalized interaction dynamics and network structure combine to shape the collective dynamics.

One promising direction for future research lies in evolutionary dynamics on temporal networks. 
Time-varying network structure is a recurring theme in social systems, encoding not only who interacts with whom but with when (and how often) these interactions happen \cite{holme2012temporal}.
It was recently discovered that temporal networks generally enhance the evolution of cooperation relative to comparable static networks \cite{li2020evolution}, yet the practical scenarios easily trigger the heterogeneous time rhythm of strategy updating.
In real temporal networks, a node's degree may vary drastically even over short time periods \cite{barabasi2005origin,karsai2018bursty,Renaud2016}.
This---in tandem with other temporal effects such as burstiness and multi-frequency interactions \cite{barabasi2005origin,karsai2021}---may lead to more exotic evolutionary dynamics.
By regarding a temporal network as a sequence of static snapshots, our theory might be adopted to further tailor individuals' update rates in temporal evolutionary game dynamics.

\section*{Appendix A: Fixation probability}
In each round of the game, individuals interact with their neighbors and accumulate the payoffs accordingly.
The payoff matrix of the game is given by
 \begin{equation}
     \bordermatrix{
         & \text{C} & \text{D} \cr
         \text{C} & b-c & -c \cr
         \text{D} & b & 0 \cr
     }.\nonumber
 \end{equation}
The state of network at any given time can be encoded by a binary vector $\mathbf{x}\in \{0,1\}^N$, where $x_i=1$ denotes that the player $i$ chooses strategy $\text{C}$, otherwise $x_i=0$ indicates strategy $\text{D}$.
Using this representation of the network state $\mathbf{x}$, $i$'s average payoff is $f_i(\mathbf{x}) = -cx_i+b\sum_{j}p_{ij} x_j$, where $p_{ij}=e_{ij}/k_i$ indicates the probability of a single step random walk from $i$ to $j$ on the network.
For a node $i$ with update rate $\lambda_i$, the probability to be chosen for a strategy update is $\lambda_i/\Lambda$.
It follows that at the end of each round, the probability for a player $j$ to transmit its strategy to $i$ is $r_{ji}(\mathbf{x}) = \frac{\lambda_i}{\Lambda} \frac{e_{ij} F_j(\mathbf{x})}{\sum_{l} e_{il}F_l(\mathbf{x})}$.

 As shown in the Supplemental Material Sec.~I \cite{Supplemental}, the fixation probability of cooperation is derived by a first-order expression as the neutral fixation probability ($1/N$) plus a correction term due to weak selection, namely
\begin{equation}
	\rho_{C}=\frac{1}{N}+\delta\left\langle\left.\frac{d}{d \delta}\right|_{\delta=0} \widehat{\Delta}(\mathbf{x})\right\rangle_{\textbf{u}}^{\circ}+O\left(\delta^{2}\right),
	\label{rhoc_general}
\end{equation}
where $\widehat{\Delta}(\mathbf{x})$ denotes the reproductive-value-weighted frequency change of cooperation, which is given by
\begin{equation}
	\widehat{\Delta}(\mathbf{x})=\sum_{i} \frac{k_i}{\lambda_i \sum_{l} \frac{k_l}{\lambda_l}} \sum_{j}\left(x_{j}-x_{i}\right) r_{j i}(\mathbf{x}).
	\label{delta_sel}
\end{equation}
Here $\left\langle \varphi  \right\rangle_{\textbf{u}}^{\circ}$ indicates the summation of the expectation of $\varphi$ under neutral drift through time step $t=0$ to infinity, namely $\langle \varphi(\mathbf{x}) \rangle_{\textbf{u}}^{\circ}=
\sum_{t=0}^{\infty} \sum_{\mathbf{x} \in \{0,1\}^N} \mathbb{P}_{\textbf{u}}^{\circ}\left[\mathbf{X}(t)=\mathbf{x} \right] \varphi(\mathbf{x})
$, where $\mathbb{P}_{\textbf{u}}^{\circ}\left[\mathbf{X}(t)=\mathbf{x} \right]$ indicates the neutral probability of the network reaching state $\mathbf{x}$ at time step $t$ starting from the initial state with a single uniformly selected cooperator in population with $N-1$ defectors.
Combining Eqs.~(\ref{rhoc_general}) and (\ref{delta_sel}), the fixation probability can be expressed as
\begin{equation}
\begin{aligned}
	\rho_{C}=&\frac{1}{N}
	+ \frac{\delta}{\Lambda \sum_{i}\frac{k_i}{\lambda_i}}\left [-c\sum_{i,j}k_i p_{ij}^{(2)}\eta_{ij} \right.\\
	&\left.+b\left(\sum_{i,j}k_i p_{ij}^{(3)}\eta_{ij}-\sum_{i,j}k_i p_{ij}\eta_{ij}\right)\right]
	+O\left(\delta^{2}\right), \nonumber
 \end{aligned}
\end{equation}
where $\eta_{ij}=\left\langle \widehat{x}-x_ix_j \right\rangle_{\textbf{u}}^{\circ}$, and $\hat{x}$ represents the reproductive-value-weighted frequency of cooperators.
Here $\eta_{ij}$ satisfies the recurrence relation of
\begin{equation}
	\eta_{ij} = \begin{cases}\frac{\Lambda}{N (\lambda_i + \lambda_j)}+\sum_{k}\frac{\lambda_i p_{ik} \eta_{kj}}{\lambda_i+\lambda_j} + \sum_{k}\frac{\lambda_j p_{jk} \eta_{ki}}{\lambda_i+\lambda_j} , & i \neq j \\ 0, &  i=j\end{cases}.
	\label{eta_ij}
\end{equation}
By letting $\rho_C>1/N$, we obtain $\mathcal{C}^*$ shown in Eq.~(\ref{exact_bcr}).

\section*{Appendix B: Calculation of the critical ratio $\mathcal{C^*}$}
We first define $\eta^{(n)} = \sum_{i,j }k_i p_{ij}^{(n)}\eta_{ij}/K$, then Eq.~(\ref{exact_bcr}) can be rewritten as 
\begin{equation}
	\mathcal{C}^* =  \frac{\eta^{(2)}}{\eta^{(3)}-\eta^{(1)}}. \nonumber
	\label{bcr_simple}
\end{equation}
From the recurrence relation of $\eta_{ij}$ in Eq.~(\ref{eta_ij}), we further derive the recurrence relation of $\eta^{(n)}$ with
\begin{equation}
	\eta^{(n)} =\sum_{i,j }\frac{k_i}{K} p_{ij}^{(n)}\frac{\Lambda}{N(\lambda_i+\lambda_j)}+\widetilde{\eta}^{(n+1)}-\sum_{i }\frac{k_i}{K} p_{ii}^{(n)}\eta_{ii}^{+},
	\label{recurr_eta}
\end{equation}
where $\widetilde{\eta}^{(n+1)}=\sum_{i,j,l }\frac{k_i}{K} p_{ij}^{(n)}\frac{2\lambda_j}{\lambda_i+\lambda_j}p_{jl}\eta_{il}$ and $\eta_{ii}^{+}=\frac{\Lambda}{2N\lambda_i}+\sum_{l}p_{il}\eta_{il}$.

By defining the difference $\Delta_{\widetilde{\eta}^{(n)}} := \widetilde{\eta}^{(n)}-\eta^{(n)}$ and using the recurrence relation of Eq.~(\ref{recurr_eta}), we obtain the calculation of $\mathcal{C^*}$ shown in Eq.~(\ref{bcr_approx}) with mean-field approximation,  with $\Delta_{\widetilde{\eta}_n} = -\Delta_{\widetilde{\eta}^{(2)}}+\frac{K^2}{\sum_{i} k_i^{2} }\Delta_{\widetilde{\eta}^{(\infty)}}$ and $\Delta_{\widetilde{\eta}_d}=-\Delta_{\widetilde{\eta}^{(2)}}-\Delta_{\widetilde{\eta}^{(3)}}+\frac{K N}{\sum_{i } k_i^{2} }\Delta_{\widetilde{\eta}^{(\infty)}}$ for simplification, where $\Delta_{\lambda^{(1)}} = \sum_{i,j }\frac{k_i}{2K} \left[1-\frac{\Lambda}{ N\lambda_i} +p_{ij}\left(1-\frac{2\Lambda}{N(\lambda_i+\lambda_j)}\right)\right]$ and $\Delta_{\lambda^{(2)}} =  \sum_{i,j }\frac{k_i}{2K} (p_{ij}+p_{ij}^{(2)})(1-\frac{2\Lambda}{N(\lambda_i+\lambda_j)})$.
According to Supplemental Material Sec.~III \cite{Supplemental}, we further have
$\Delta_{\widetilde{\eta}^{(2)}}\approx N\Delta_{\widetilde{\eta}^{(\infty)}}/\langle k \rangle $ and  $\Delta_{\widetilde{\eta}^{(3)}} \approx N\Delta_{\widetilde{\eta}^{(\infty)}}/\langle k \rangle^2$ for large networks, and hence $\mathcal{C^*}$ shown in Eq.~(\ref{bcr_approx_heter_large}) follows immediately.

\end{document}